# Electron paramagnetic resonance signatures of $Co^{2+}$ and $Cu^{2+}$ in $\beta$-$Ga_2O_3$

Jan E. Stehr[1]*, Detlev M. Hofmann[2], Weimin M. Chen[1], and Irina A. Buyanova[1]‡

[1] Department of Physics, Chemistry and Biology, Linköping University, 58183 Linköping, Sweden

[2] Physikalisches Institut, Justus-Liebig-University Giessen, Heinrich-Buff-Ring 16, 35392 Giessen, Germany

Corresponding authors:
*Jan E. Stehr (jan.eric.stehr@liu.se)

‡ Irina A. Buyanova (irina.bouianova@liu.se)

**ABSTRACT**

Gallium oxide ($\beta$-$Ga_2O_3$) is a wide-bandgap compound semiconductor with a bandgap of ~ 4.9 eV that is currently considered promising for a wide range of applications ranging from transparent conducting electrodes to UV optoelectronic devices and power electronics. However, all of these applications require a reliable and precise control of electrical and optical properties of the material, which can be largely affected by impurities, such as transition metals commonly present during the growth. In this work we employ electron paramagnetic resonance (EPR) spectroscopy to obtain EPR signatures of the 3d-transition metals $Co^{2+}$ and $Cu^{2+}$ in $\beta$-$Ga_2O_3$ bulk crystals and powders that were unknown so far. Furthermore, we show that $Co^{2+}$ and $Cu^{2+}$ both preferentially reside on the octahedral gallium lattice site.

β-Ga$_2$O$_3$ is a wide bandgap semiconductor that is attractive for various applications [1], including in power electronics [2,3], as transparent conductive electrodes [4], in solar-blind UV photodetectors [5] and gas sensors [6], as well as for photoelectrochemical water splitting [7]. For example in the case of high-power electronics, β-Ga$_2$O$_3$ is predicted to surpass the current state-of-the-art technology based on GaN and SiC, due to a higher breakdown field [1]. Another significant advantage of β-Ga$_2$O$_3$ is that large bulk crystals can be grown by melt growth techniques, giving access to high quality and reasonably priced native substrates, which is essential for the fabrication of high-performance power devices. The key to the realization of device applications is to achieve control over conductivity by doping and mitigation of trap states, as most of the electronic properties of β-Ga$_2$O$_3$ are affected by the presence of dopants/contaminants and/or intrinsic defects. Here, transition metals (TMs) represent an important group of impurities, which are either unintentionally present during the growth or are used as intentional dopants. These elements mostly introduce deep-level states in β-Ga$_2$O$_3$ limiting its conductivity [1,8]. For example, Fe is used as the main compensating dopant to fabricate semi-insulating β-Ga$_2$O$_3$. On the other hand, incorporation of transition metals, e.g. Co and Ni, can also significantly enhance photocatalytic properties of β-Ga$_2$O$_3$ [9]. Moreover, Mn- and Fe-doped Ga$_2$O$_3$ have been shown to exhibit room-temperature ferromagnetism [10,11], promising for room temperature spintronics. From this point of view Cu is also of special interest since it was calculated that Cu-doped β-Ga$_2$O$_3$ has 100% spin polarization of states near the Fermi level, which makes it very attractive for spintronic applications [12]. Furthermore, it was predicted that Cu dopants form shallow acceptor levels, indicating that Cu doping might lead to p-type β-Ga$_2$O$_3$ [12]. Thus, it is of crucial interest to understand the electronic structure of TMs and their interactions with intrinsic defects and impurities. Though spectroscopic signatures of several TMs, such as Fe, Cr, Mn and Ti, have recently been obtained from electron paramagnetic resonance (EPR) studies [13–16], electronic structure of other commonly present TM, including Co and Cu, remains unknown so far. This has motivated the present study of TM signatures in undoped and cobalt doped β-Ga$_2$O$_3$ by employing EPR, as this technique is known to be among the most powerful and versatile experimental methods for identification of defects and impurities [17].

We used commercially available β-Ga$_2$O$_3$ powder from Sigma Aldrich and β-Ga$_2$O$_3$:Co bulk crystals grown by the Czochralski method [18]. The β-Ga$_2$O$_3$:Co sample has a Co$^{2+}$ concentration of $(10\pm5) \times 10^{17}$ cm$^{-3}$ determined by EPR spin counting. For the β-Ga$_2$O$_3$ powder we estimated the Cu$^{2+}$ concentration to be $\sim 5 \times 10^{15}$ cm$^{-3}$. The samples were measured in a X-band resonator of a Bruker E500 spectrometer equipped with a He-gas flow cryostat for measurements with adjustable temperatures ranging from 5 K to 300 K in the dark. EPR spectra were analyzed using the following spin-Hamiltonian that includes an electron Zeeman term and a central hyperfine interaction term:

$$\mathcal{H} = \mu_B \boldsymbol{BgS} + \boldsymbol{SAI} \qquad (1)$$

Here, **S** denotes the effective electron spin, **I** the nuclear spin and B is an external magnetic field. **g** and **A** are the electron g-tensor and the hyperfine interaction tensor, respectively and $\mu_B$ is the Bohr magneton. Modeling of the EPR spectra was done using the Easyspin software package [19].

Figure 1 depicts an EPR spectrum of a β-Ga$_2$O$_3$ bulk crystal measured at 6 K with an orientation of the external magnetic field parallel to the a*-axis of the crystal. The signal is centered around 105 mT and consists of one group of 8 equidistant lines (marked by the black rake) indicating that it stems from a paramagnetic center with an electron spin S = ½ and a resolved hyperfine interaction involving a nuclear spin I = 7/2 with 100 % natural abundance. In order to determine the full set of spin-Hamiltonian parameters given in Eq. 1, angular dependent EPR measurements were performed.

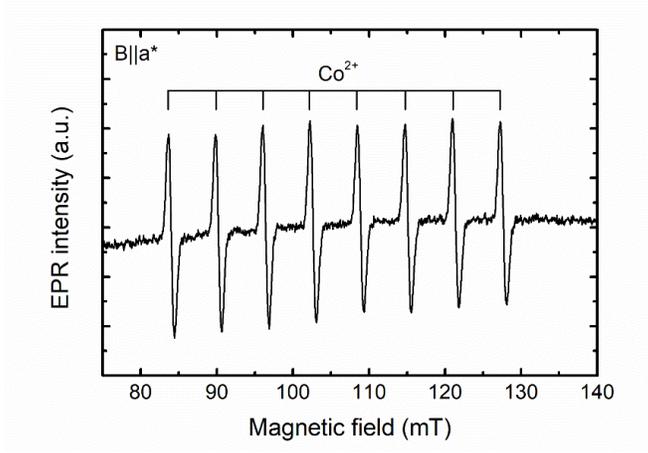

*Figure 1: EPR spectrum of $Co^{2+}$ in $\beta$-$Ga_2O_3$ measured at 6 K with $B \parallel a^*$.*

Figure 2 shows EPR peak positions measured at 6 K as a function of the angle θ between **B** and the a*-axis for rotations around the crystallographic [010] axis (c) and the [001] axis (d). The crystal structure of $\beta$-$Ga_2O_3$ is displayed in figure 2 (a). By fitting the experimental data given by the open circles by Eq. 1 we can determine the g-values as $g_{a^*}$ = 6.34±0.01, $g_b$ = 3.37±0.01 and $g_c$ = 2.0±0.01, while the hyperfine interaction tensor is given by $|A_{a^*}|$ = (550±10) MHz , $|A_b|$ = (60±5) MHz and $|A_c|$ = (130±10) MHz. Here the subscripts denote components of the g- and A-tensors that are parallel to the given crystallographic axes.

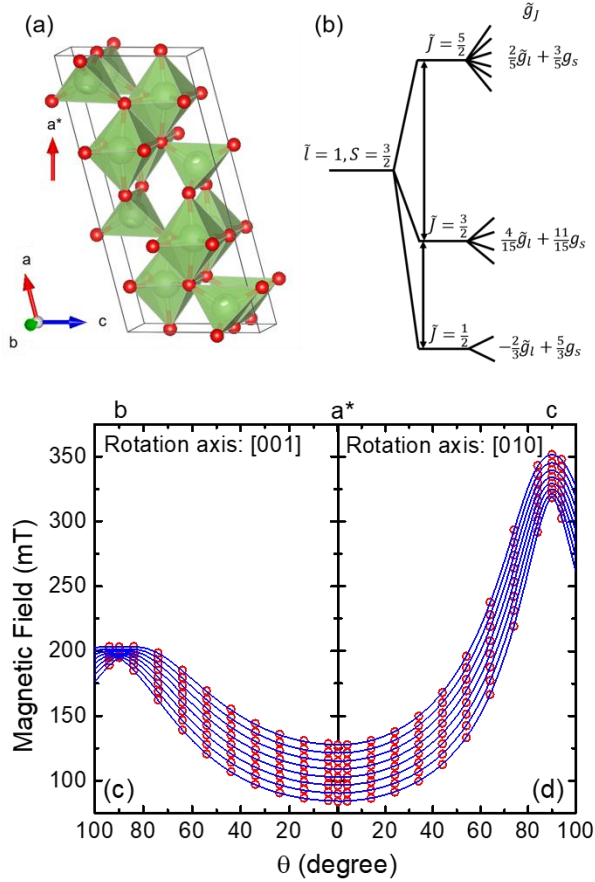

*Figure 2: (a) The lattice and the main crystallographic axes of β-Ga$_2$O$_3$ (b) electronic structure of Co$^{2+}$. EPR resonance field positions of Co$^{2+}$ in β-Ga$_2$O$_3$ measured at 6 K as a function of the angle θ for rotation around the crystallographic [010] axis (c) and the [001] axis (d) i.e. the a-axis and b-axis as defined in (a). The experimental data are shown by the open circles, while the simulation results using the spin-Hamiltonian in Eq.1 are depicted by the solid lines.*

Let us now discuss the origin of this EPR signal. The only likely elements, which fulfill the criteria of the observed hyperfine interaction with a nucleus with I = 7/2 and 100% natural abundance, are scandium ($^{45}$Sc), vanadium ($^{51}$V), cobalt ($^{59}$Co), holmium ($^{165}$Ho) and tantalum ($^{181}$Ta). Sc$^{2+}$ has a 3d$^1$ electron configuration and g-values of $g_\parallel = 1.94$ and $g_\perp = 1.98$ can be found in the literature [20]. These values do not fit the experimental data and, therefore, scandium can be excluded as the chemical origin of the observed EPR signal. Vanadium has been observed by EPR in three different charge states, V$^{2+}$ in a 3d$^3$ electron configuration, V$^{3+}$ in a 3d² electron configuration and $V^{4+}$ in a 3d¹ electron configuration. However, $V^{3+}$ can be excluded since it has the electron spin of S=1, while $V^{2+}$ and $V^{4+}$ can be excluded due to their anisotropy and g-tensors of g ~ 2 and g ~ 1.2, respectively [20,21]. $Ho^{2+}$ and $Ho^{3+}$ have both a very large hyperfine splitting and g-values [20], as observed by Boyn et al. [22] and Shakurov et al. [23], which differs significantly from the parameters observed for this center. This also excludes Ho as the chemical origin of the observed EPR signal. In the case of Ta only little information on EPR data is available. Irmscher et al. observed an EPR center in 6H-SiC that they labeled as $Ta^{3+}$ with S=1 that showed a g-value around 2 and has a very large nuclear quadrupole moment [24]. $Ta^{4+}$ was observed in $TiO_2$ with S = ½ and an almost isotropic g-tensor of 2 [25]. Since the signal observed here differs significantly from both reported cases for Ta, we can also exclude it.

Thus, the last element to consider is cobalt. $Co^{2+}$ has a (3d⁷) ⁴F electronic ground state configuration. If one assumes a purely octahedral crystal field, the lowest orbital state is a triplet ($\Gamma_4$), which is split by spin-orbit coupling into three orbitally-degenerate states, each of which is fourfold spin-degenerate. In zero magnetic field, these 12 levels are expected [26] to split into a doublet, a quadruplet and a sextet, among which the doublet is the lowest in energy as schematically depicted in figure 2 (b). Spin resonance can be only observed for the lowest doublet, resulting in an effective electron spin of $S_{eff}$ = 1/2. The evaluation of the Zeeman effect within this doublet using the spin-Hamiltonian operator $\tilde{g}_l l + g_s S$ yields the isotropic $\tilde{g}$-factor [20]:

$$\tilde{g} = \frac{5}{3}g_s - \frac{2}{3}\tilde{g}_l \qquad (2)$$

Here $g_s$ ($\tilde{g}_l$) are the free electron spin (effective orbital) g-factor. With $g_s = 2$ and $\tilde{g}_l = -\frac{3}{2}$ this results in $\tilde{g} \sim 4.3$ [27]. Due to the presence of a large orbital angular momentum in the $\Gamma_4$ triplet, a large deviation of the ground-state g-factor from the free electron value of 2.0023 can be observed. However, most cobaltous salts show very high anisotropy: in the case of an octahedral coordination the g-values are usually close to $g_\perp = 2.95$ and $g_\parallel = 6.24$ (see Abragam and Bleaney [20] or Pilbrow [27]). Here $g_\parallel$ refers to an orientation of an external magnetic field parallel to an axis going through the top and bottom of the octahedron, which coincides with the a*-axis in this case, while $g_\perp$ refers to $g_b$. This anisotropy is caused by small trigonal or tetragonal distortions of the octahedron. One can describe this effect by adding terms to the energy matrices, which results in a splitting of the quadruplet and sextet. The parallel and perpendicular g-values are then given by [27]:

$$g_\parallel = \frac{5}{3}g_s - \frac{2}{3}\tilde{g}_l + \left(\frac{4\sqrt{5}a}{3}\right)(2g_s - \tilde{g}_l) \tag{3}$$

$$g_\perp = \frac{5}{3}g_s - \frac{2}{3}\tilde{g}_l - \left(\frac{2\sqrt{5}a}{3}\right)(2g_s - \tilde{g}_l) \tag{4}$$

The effective orbital g-factor $\tilde{g}_l$ equals to -3/2 for the triplet orbital ground state with a fictitious angular momentum $\tilde{l} = 1$. The parameter $a$ is a measure of the distortion and is small as compared to unity. From the experimental $g_\parallel$-value, the distortion parameter $a$ is calculated to be 0.12 using Eq. 3 and 4. The obtained value is in agreement with the values for several cobaltous salts [20]. The g-factors roughly follow the relation:

$$g_\parallel + 2g_\perp \approx 5g_s - 2\tilde{g}_l \approx 13 \tag{5}$$

In our case the obtained value of 13.1 is indeed very close to the model. Cobalt on the tetrahedral gallium site is less likely as it is expected to have nearly isotropic g-values around 2.4 [20].

Therefore, the observed EPR signal can be assigned to $Co^{2+}$ located at the octahedral gallium site (0.62 Å lattice space). This position seems to be preferable as compared to the tetrahedral gallium site (0.477 Å lattice space), since it provides more space for the $Co^{2+}$ ion (0.75 Å ion radius) which has a larger ion radius as compared with the $Ga^{3+}$ ion with the radius of 0.62 Å. Also, the observed spin-Hamiltonian parameters are quite similar to those observed for $Co^{2+}$ on an octahedral lattice site in $\alpha$-$Al_2O_3$ [28].

In the case of $\beta$-$Ga_2O_3$ powder, the EPR spectrum is found to be rather different as can be seen from figure 3 (a). It now contains two groups of four EPR lines (indicated by the black rakes) located at 280-304 and 320-350 mT, i.e. at g-values in the vicinity of 2.32 and 2.02. These g-values are characteristic for acceptor centers. The four lines are caused by a resolved hyperfine interaction with a nuclear spin $I = 3/2$. A close examination of the low field lines shows that they also contain a weaker set of four lines (indicated by the blue rake), which cannot be resolved in case of the second group at $g = 2.02$ due to low intensity and strong overlap with the more intense set. This suggests that the involved paramagnetic defect has two isotopes with the same nuclear spin $I = 3/2$. By comparing the signal intensities from both groups, the isotope ratio can be estimated. The calculation yields 70 % (30%) for the signals marked by a black (blue) rake. The ratio of the line spacing within these sets implies that the ratio of nuclear magnetic moments between the two isotopes is 1.077. Figure 3 (b) shows a simulated EPR powder spectrum of a defect with $I = 3/2$ with the following simulation parameters:

70% abundance: $g_{\parallel} = 2.33 \pm 0.01$, $g_{\perp} = 2.02 \pm 0.01$, $|A| = (195 \pm 10)\ MHz$

30% abundance: $g_{\parallel} = 2.31 \pm 0.01$, $g_{\perp} = 2.02 \pm 0.01$, $|A| = (210 \pm 10)\ MHz$

The simulations are in excellent agreement with the experimental data, justifying the proposed model.

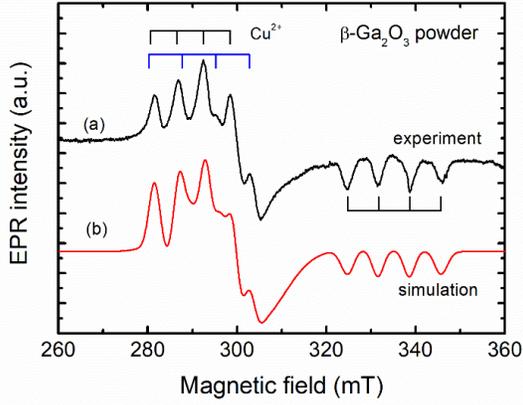

Figure 3: (a) Experimental EPR spectrum of $Cu^{2+}$ in $\beta$-$Ga_2O_3$ powder measured at 15 K. (b) Simulated EPR spectrum of $Cu^{2+}$ in $\beta$-$Ga_2O_3$ powder with the spin-Hamiltonian parameters given in the text.

Cu is the only element in the periodic table fulfilling the above criteria with the two isotopes $^{63}$Cu (69.2 % natural abundance) and $^{65}$Cu (30.8 % natural abundance). These isotopic abundances are in excellent agreement with the experimentally determined isotope ratio (70 % and 30%). Also, the ratio of the observed hyperfine interaction $\frac{A(^{65}Cu)}{A(^{63}Cu)} = \frac{210\ MHz}{195\ MHz} = 1.077$ is in excellent agreement with that of their nuclear magnetic moments $\frac{\mu(^{65}Cu)}{\mu(^{63}Cu)} = \frac{2.38}{2.22} = 1.072$ [29].

There are three possible charge states for Cu located on a gallium lattice site, namely, $Cu^+$ with a $3d^{10}$ electron configuration and $S = 0$, $Cu^{2+}$ with a $3d^9$ electron configuration and $S = 1/2$ and $Cu^{3+}$ with a $3d^8$ electron configuration and $S = 1$. Since the detected EPR signal stems from the paramagnetic center with $S = 1/2$, it must be related to $Cu^{2+}$. One possible location for the $Cu^{2+}$ ion is the tetrahedral gallium position. However, in this case the g-values should be similar to the g-values of $Cu^{2+}$ in GaN [30] or in ZnO [31] ($g_\parallel \leq 0.7$ and $g_\perp \sim 1.5$) that are very different from the g-values observed for the center here. On the other hand, $Cu^{2+}$ located at the octahedral lattice site in other materials shows g-values ($g_\parallel \sim 2.4$ and $g_\perp \sim 2.1$) very similar to the ones obtained from the experimental data, see e.g. Abragam and Bleaney ($3d^9$ $Cu^{2+}$ in an octahedral field) [20] or Keeble et al. ($Cu^{2+}$ in $PbTiO_3$) [32]. Thus, the observed signal can be assigned to $Cu^{2+}$ located on the octahedral gallium site.

In conclusion, we have employed EPR spectroscopy to investigate the electronic structure and geometric arrangement of cobalt and copper in $\beta$-$Ga_2O_3$. We show that both of Co and Cu are present in undoped $\beta$-$Ga_2O_3$ in the 2+ charge state in the $3d^7$ and $3d^9$ electronic configuration, respectively. Detailed angular-dependent EPR measurements yielded accurate spin Hamiltonian parameters, such as g-tensor and the hyperfine interaction tensor, of the $Co^{2+}$ and $Cu^{2+}$ centres. The obtained parameters provide signatures of these TMs that can be used for their identification. Owing to their large ionic radii, both of these impurities preferentially occupy the octahedral Ga (II) lattice site, i.e. having the same lattice configuration as other TMs in $\beta$-$Ga_2O_3$.

Acknowledgements.

We would like to thank Leibniz-Institut für Kristallzüchtung (Berlin) for providing the $\beta$-$Ga_2O_3$ bulk crystal sample. Financial support by Linköping University through the Professor Contracts and the Swedish Government Strategic Research Area in Materials Science on Functional Materials at Linköping University (Faculty Grant SFO-Mat-LiU No 2009 00971) is greatly appreciated.